# Ejby—A new H5/6 ordinary chondrite fall in Copenhagen, Denmark

H. HAACK [1]*, A. N. SØRENSEN[2], A. BISCHOFF[3], M. PATZEK [3], J.-A. BARRAT[4], S. MIDTSKOGEN[5], E. STEMPELS[6], M. LAUBENSTEIN [7], R. GREENWOOD[8], P. SCHMITT-KOPPLIN[9,10], H. BUSEMANN[11], C. MADEN[11], K. BAUER[11], P. MORINO[11], M. SCHÖNBÄCHLER[11], P. VOSS[12], and T. DAHL-JENSEN[12]

[1]Maine Mineral and Gem Museum, 99 Main St., Bethel, Maine 04217, USA
[2]The Niels Bohr Institute, University of Copenhagen, Blegdamsvej 17, 2100 Copenhagen Ø, Denmark
[3]Institut für Planetologie, Westfälische Wilhelms-Universität Münster, Wilhelm-Klemm-Str. 10, 48149 Münster, Germany
[4]Université de Bretagne Occidentale, Institut Universitaire Europe0en de la Mer, CNRS, UMR 6538, Place Nicolas Copernic,29280 Plouzané, France
[5]The Norwegian Meteor Network, Harestua, Norway
[6]Department of Physics & Astronomy, Uppsala University, Box 516, SE-75210 Uppsala, Sweden
[7]Istituto Nazionale di Fisica Nucleare—Laboratori Nazionali del Gran Sasso, Via G. Acitelli 22, 67100 Assergi, AQ, Italy
[8]Planetary and Space Sciences, School of Physical Sciences, The Open University, Walton Hall, Milton Keynes, MK7 6AA, UK
[9]Research Unit Analytical BioGeoChemistry (BGC), Helmholtz Zentrum München, Ingolstädter Landstrasse 1, 85764 Neuherberg, Germany
[10]Chair of Analytical Food Chemistry, Technische Universität München, Alte Akademie 10, 85354 Freising, Germany
[11]Inst. für Geochemie und Petrologie, ETH Zurich, Clausiusstrasse 25, 8092 Zürich, Switzerland
[12]Geological Survey of Denmark and Greenland, Øster Voldgade 10, 1350 Copenhagen K, Denmark
Corresponding author. E-mail: hhaack@mainemineralmuseum.org

**Abstract**– On February 6, 2016 at 21:07:19 UT, a very bright fireball was seen over the eastern part of Denmark. The weather was cloudy over eastern Denmark, but many people saw the sky light up-even in the heavily illuminated Copenhagen. Two hundred and thirty three reports of the associated sound and light phenomena were received by the Danish fireball network. We have formed a consortium to describe the meteorite and the circumstances of the fall and the results are presented in this paper. The first fragment of the meteorite was found the day after the fall, and in the following weeks, a total of 11 fragments with a total weight of 8982 g were found. The meteorite is an unbrecciated, weakly shocked (S2), ordinary H chondrite of petrologic type 5/6 (Bouvier et al. 2017). The concentration of the cosmogenic radionuclides suggests that the preatmospheric radius was rather small ~20 cm. The cosmic ray exposure age of Ejby (83 ± 11 Ma) is the highest of an H chondrite and the second highest age for an ordinary chondrite. Using the preatmospheric orbit of the Ejby meteoroid (Spurny et al. 2017) locations of the recovered fragments, and wind data from the date of the fall, we have modeled the dark flight (below 18 km) of the fragments. The recovery location of the largest fragment can only be explained if aerodynamic effects during the dark flight phase are included. The recovery location of all other fragments are consistent with the dark flight modeling.

## INTRODUCTION

On February 6, 2016 at 21:07:19 UT, a very bright fireball was observed over eastern Denmark. The local time was 22:07 and since it was a Saturday night, a lot of people were outdoors and despite the cloud cover, they saw the skies light up over Copenhagen and eastern Zealand. The skies were clear over Germany, Austria, and on the Danish West coast, and fortunately, the event was caught on several cameras. Three still images from Germany and Austria, a high-resolution radiometric light curve from the Czech Republic, and a video from the Danish west coast recorded the fireball (Spurny et al. 2017). Two seismic stations in Copenhagen also detected a signal with an N-wave shape, characteristic of supersonic shock waves. Following extensive media coverage of the event, many people searched for meteorites the following day a Sunday with nice weather. A single 46-g fragment was found on the first day in Ejby, which is a suburb 10 km WNW of the center of Copenhagen. The fragment had hit the tiles outside the front door of a private house. A piece of the otherwise completely crusted meteorite had broken off; however, after a search along the house, we (HH and his son) found two of the missing pieces weighing 2 and 10 g. During the following weeks, another 10 fragments were found giving a total recovered weight of 8982 g, making Ejby the biggest recorded meteorite fall in Denmark (Table 1). Most of the fragments recovered after Monday were wet, due to heavy rain on Monday morning. Several fragments lost weight after they were brought into the museum to dry up. Fragments with nearly complete fusion crust lost minimal weight, presumably because the fusion crust protected them from the rain. The biggest fragment, described as the Herlev fragment in the following text, hit a tiled courtyard in an industrial area in Herlev, where it was found by the owner Monday morning. The fragment had shattered into countless pieces which were found scattered across the courtyard. All of these fragments were soaked in rain and had rusty stains already. Four large (350 - 510 g) pieces were found close to the first fragment in Ejby. All of these had a fractured surface without fusion crust. Two of them had penetrated the ground to a depth of 10–20 cm and we originally thought that they had broken upon impact. We carefully searched the ground within and around the holes, but did not find any additional pieces. Another surprising finding is that the fragments were unevenly distributed within the strewn field (Fig. 1). Since the entire strewn field is within a densely populated area, within greater Copenhagen, the chance of finding a meteorite should be high and approximately the same anywhere within the strewn field. A large cluster of fragments in the suburb Ejby suggest that the last fragmentation event was a relatively low altitude above Ejby. This is consistent with the observation that four of the pieces had fractured surfaces without fusion crust. Two of the Danish seismic stations recorded the supersonic boom at 21:08:1.23 and 21:08:3.93 (UTC). Since the largest fragment hit a tiled surface only 6.5 km from the closest seismic station, we looked for a seismic signal from the impact but did not find any. As required by the Danish "Danekræ" law of 1991, all of the recovered fragments were handed over to the Natural History Museum of Denmark. The finders have received a finders' fee which is based on an evaluation of the market values of the recovered specimens.

Following the recovery of the meteorites, we have formed a consortium to study the meteorite. In this paper, we present data on the darkflight modeling, mineralogy, chemistry, Ti isotope cosmochemistry, cosmic ray exposure (CRE) history, organics, and noble gases in the new meteorite.

Table 1. Masses and locations of the recovered Ejby meteorites.

| Sample No | Mass (g) | Coordinates | | Suburb | NHMD Sample no. |
|---|---|---|---|---|---|
| 1 | 58.8 | 55°42′.05″N | 12°24′19.92″E | Ejby | 2019.10 |
| 2 | 319.0 | 55°42′5.01″N | 12°24′48.47″E | Ejby | 2019.11 |
| 3 | 6695.8 | 55°42′59.78″N | 12°26′59.59″E | Herlev | 2019.12 |
| 4 | 64.6 | 55°40′53.10″N | 12°24′50.52″E | Glostrup | 2019.13 |
| 5 | 58.5 | 55°41′34.27″N | 12°25′59.52″E | Rødovre | 2019.14 |
| 6 | 510.4 | 55°42′16.00″N | 12°24′24.00″E | Ejby | 2019.15 |
| 7 | 437.7 | 55°42′20.00″N | 12°24′7.00″E | Ejby | 2019.16 |
| 8 | 431.0 | 55°42′21.17″N | 12°24′13.82″E | Ejby | 2019.17 |
| 9 | 350.0 | 55°42′18.37″N | 12°24′11.97″E | Ejby | 2019.18 |
| 10 | 38.5 | 55°42′17.47″N | 12°24′34.79″E | Ejby | 2019.19 |
| 11 | 18.1 | 55°41′44.1″N | 12°28′40.2″E | Vanløse | 2019.20 |
| Total mass | 8982.4 | | | | |

## ANALYTICAL PROCEDURES

### Dark-Flight Modeling

The luminous parts of the trajectory were first observed at height of 85.5 km and could be followed down to a height of 18.3 km. The heading was 44° and the slope was 62° (Spurny et al. 2017). In order to recover and model the nonluminous atmospheric flight of the meteoroid, including effects such as wind drift, we have performed so-called dark-flight modeling. Dark-flight modeling was performed using our Python based code that numerically solves the movement of a meteoroid through the Earth's atmosphere, given the preatmospheric mass, the efficiency of ablation, the structure of the atmosphere, and the distribution of wind as a function of altitude. The interaction of the atmosphere on the dark-flight trajectory is calculated by solving the equations from Pecina and Ceplecha (1983) as an initial value problem. We have also implemented nonspherical shapes for the meteoroid in a similar fashion as Gritsevich (2009), by introducing a relationship between the meteoroid density and shape in terms of a shape-factor A, and the orientation/rotation during flight in terms of the shape-change parameter l. Fragmentation is modeled as an instantaneous disintegration of the meteoroid into smaller fragments.

Unfortunately, the obtained video data are sparse, and although Bartoschewitz et al. (2017) and Spurny et al. (2017) were able to estimate the direction and preatmospheric velocity (14.52 _ 0.10 km s _1) of the meteoroid, it was not possible to measure the of the meteoroid in the final, lowest part of the luminous trajectory. A proper estimate of the deceleration is normally needed to obtain an estimate of the preatmospheric mass and ablation coefficient. However, in this case, an educated guess of the original mass is possible by combining the total mass of the recovered meteorite, its density, and the radiometric estimate of the meteoroid radius. From this, we estimate the preatmospheric mass of Ejby to be around 120 kg. This is consistent with the size estimate based on cosmogenic isotopes which suggest preatmospheric radius of less than 20 cm. A spherical mass with a radius of 20 cm and a density of 3300 kg m$^{-3}$ has a mass of 110 kg, but a more irregular mass with a maximum shielding depth of 20 cm would be heavier. Other parameters used in the dark flight modeling are a density of 3300 kg m$^{-3}$, a conservative ablation factor of r = 0.020 s$^2$ km$^{-2}$, and random rotation (l = 2/3, but see also further below). The atmospheric drag coefficient cd is estimated according to the description of Carter et al. (2009). This drag coefficient is calculated for a spherical body; in order to account for nonsphericity, the shape-factor was assumed to be A = 1.8, following the results of Gritsevich (2009). Wind and pressure profiles were obtained from the Danish Meteorological Institute.

The still images and the radiometric light curve indicate several flaring events and, thus, fragmentation (Bartoschewitz et al. 2017; Spurny et al. 2017). The first fragmentation happened at an altitude of 53 km, and maximum dynamic pressure occurred around 25 km. The luminous flight can be traced down to an altitude of only 18.3 km. To estimate the effect of fragmentation on the possible impact locations, we calculated darkflight trajectories for four possible fragmentation heights (53, 25, 18, and 15 km), varying the masses after fragmentation.

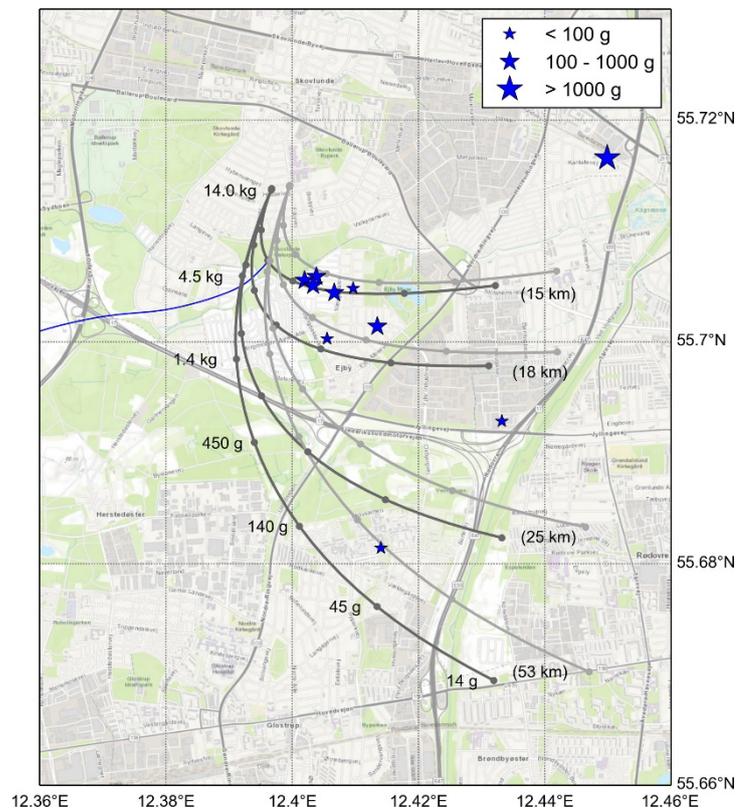

Figure 1: Distribution of the recovered fragments in the fall area 10 km WSW of central Copenhagen. The sizes of the symbols are proportional to the mass of the recovered fragments. The fireball was heading NE which is consistent with the fact that the, by far, largest fragment was found farthest to the NE. High-altitude winds were blowing from the west at the day of the fall and all fragments were therefore found east of the ground trajectory. Black and gray lines show the modeled distribution of meteorites from fragmentation events at 53, 25, 18, and 15 km. Strong, high-altitude winds were blowing from the west at the day of the fall, pushing low-mass fragments to the east. The black lines show results of modeling using the wind speeds reported by the Danish Meteorological Institute, whereas the gray lines use a 10% higher wind speed. Round symbols along the individual lines show the distribution of fragments ranging in size from 14 kg to 14 g. The blue line indicates the ground track for a fragment with a final mass of 1.4 kg, created in a fragmentation event at an altitude of 15 km.

**Mineralogy**

Several thin sections of Ejby were studied by optical and electron microscopy. The experimental procedure is described in supporting information.

**Oxygen Isotopes**

Oxygen isotope analysis was carried out at the Open University using an infrared laser-assisted fluorination system (Miller et al. 1999; Greenwood et al. 2017). An approximately 100-mg whole-rock chip of the Ejby chondrite was crushed and homogenized, and from this, two (~2 mg) replicates were extracted for oxygen isotope analysis (Table 2; see Fig. 3). The experimental procedure is described in supporting information. Oxygen isotopic analyses are reported in standard d notation, where $\delta^{18}O$ has been calculated as: $\delta^{18}O = ([^{18}O/^{16}O]_{sample}/[^{18}O/^{16}O]_{VSMOW}-1)\ 1000$ (‰) and similarly for $\delta^{17}O$ using the $^{17}O/^{16}O$ ratio. $\Delta^{17}O$, which represents the deviation from the terrestrial fractionation line, has been calculated using a linearized format (Miller 2002):

$$\Delta^{17}O = 1000\ \ln(1+ \delta^{17}O/1000) - \lambda 1000\ln(1 + \delta^{18}O/1000)$$

where $\lambda = 0.5247$, which was determined using 47 terrestrial whole-rock and mineral separate samples (Miller et al. 1999; Miller 2002).

Table 2. Oxygen isotope composition Ejby. The sample was obtained from fragment 3.

| Sample | $\delta^{17}O$‰ | 2σ | $\delta^{18}O$‰ | 2σ | $\Delta^{17}O$‰ | 2σ |
|---|---|---|---|---|---|---|
| Ejby 1 | 3.041 | | 4.400 | | 0.733 | |
| Ejby 2 | 3.035 | | 4.332 | | 0.762 | |
| MEAN | 3.038 | 0.008 | 4.366 | 0.096 | 0.748 | 0.042 |

$\Delta^{17}O$ values calculated using the linearized format of Miller (2002).

**Titanium and Chromium Isotope Analyses**

The Ti and Cr isotope data were obtained at the Institute of Geochemistry and Petrology at ETH Zürich. The analytical procedure followed that described in Bischoff et al. (2017). The experimental procedure is described in supporting information.

**Chemistry**

The chemical composition of the bulk sample was obtained using inductively coupled plasma–atomic emission spectroscopy (ICP-AES) and inductively coupled plasma–sector field mass spectrometry (ICPSFMS) (Table 3). A 1-g whole-rock sample was analyzed for major and trace element concentrations following the procedures of Barrat et al. (2012, 2016).

Table 3. Bulk chemistry of Ejby. Oxides in wt%, all others in ppm wt. Sample obtained from fragment 3.

| | ICP-AES | ICP-SFMS | | ICP-SFMS |
|---|---|---|---|---|
| $TiO_2$ | 0.11 | 0.0970 | Zr | 4.63 |
| $Al_2O_3$ | 2.07 | | Nb | 0.381 |
| $Fe_2O_3$ | 41.06 | | Cs | 0.0178 |
| MnO | 0.32 | 0.293 | Ba | 3.49 |
| MgO | 24.85 | | La | 0.344 |
| CaO | 1.69 | 1.90 | Ce | 0.885 |
| $Na_2O$ | 0.90 | | Pr | 0.132 |
| $K_2O$ | 0.08 | 0.106 | Nd | 0.658 |
| $P_2O_5$ | 0.30 | 0.304 | Sm | 0.211 |
| Ppm | | | Eu | 0.0721 |
| Cr | 3118 | | Gd | 0.303 |
| Li | | 2.21 | Tb | 0.0548 |
| Be | | 0.024 | Dy | 0.369 |
| K | | 950 | Ho | 0.0815 |
| Sc | 8 | 9.33 | Er | 0.239 |
| V | 67 | 80.6 | Tm | 0.0356 |
| Mn | | 2606 | Yb | 0.228 |
| Co | 895 | 536 | Lu | 0.0345 |
| Ni | 21,130 | | Hf | 0.135 |
| Cu | | 73.1 | Ta | 0.0107 |
| Zn | | 43.54 | W | 0.24 |
| Ga | | 5.06 | Pb | 0.057 |
| Rb | | 2.88 | Th | 0.0378 |
| Sr | 11 | 10.88 | U | 0.010 |
| Y | 2.1 | 2.97 | | |

**Noble Gases**

Noble gases were measured at ETH Zurich with standard procedures (Riebe et al. 2017) in two bulk fragments ("small," 15.41 ±0.01 mg and "large," 117.82±0.01 mg) of the Herlev fragment (see Table 1). The experimental procedure is described in supporting information.

**Soluble Organics Chemistry**

The extracts for negative mode electrospray Fourier transform ion cyclotron resonance mass spectrometry (ESI(-)-FT-ICR-MS) analysis were prepared and analyzed in the same conditions as described previously in Schmitt-Kopplin et al. (2010). The experimental procedure is described in supporting information.

**Cosmogenic Radionuclides**

Cosmogenic radionuclide concentrations have been analyzed by means of nondestructive high purity germanium gamma spectroscopy. The counting efficiencies have been calculated using thoroughly tested Monte Carlo codes. All specimens were measured in the underground laboratories at the Laboratori Nazionali del Gran Sasso of the Istituto Nazionale di Fisica Nucleare (INFN-LNGS) (Arpesella 1996; Laubenstein 2017). The specimens were received at INFN-LNGS only 1 week after the fall, so that also rather short-lived radionuclides ($T_{1/2}$ = 6–28 d, Table 4) such as $^{52}$Mn, $^{48}$V, and $^{51}$Cr could be measured.

Table 4. Massic activities (corrected to the fall of the meteorite February 6, 2016) of cosmogenic radionuclides (dpm kg$^{-1}$) in eight specimens of the Ejby meteorite measured by nondestructive gamma-ray spectroscopy. Errors include a 1σ uncertainty of 10% in the detector efficiency calibration.

| Nuclide | Half-life | Ejby #1 (46.19 g) | #2 (314.25 g) | #3a (200.18 g) | #3b (85.58 g) | #3c (82.57 g) | #4 (64.59 g) | #5 (58.5 g) | #11 (18.1 g) |
|---|---|---|---|---|---|---|---|---|---|
| $^{52}$Mn | 5.591 d | 21 ± 4 | <25 | 15 ± 4 | 16 ± 3 | 22 ± 3 | 21 ± 3 | <58 | <91 |
| $^{48}$V | 15.9735 d | 25 ± 3 | 21 ± 2 | 23 ± 2 | 19 ± 2 | 26 ± 2 | 19 ± 2 | 20 ± 2 | 22 ± 4 |
| $^{51}$Cr | 27.704 d | 52 ± 14 | 64 ± 8 | 52 ± 9 | 64 ± 9 | 73 ± 10 | 55 ± 9 | 68 ± 11 | 51 ± 13 |
| $^{7}$Be | 53.22 d | 55 ± 12 | 51 ± 6 | 54 ± 7 | 63 ± 8 | 50 ± 7 | 68 ± 9 | 88 ± 11 | 69 ± 12 |
| $^{58}$Co | 70.83 d | 7.5 ± 1.5 | 7.6 ± 0.8 | 9.4 ± 1.1 | 9.8 ± 1.2 | 8.2 ± 1.0 | 6.8 ± 0.9 | 6.6 ± 0.9 | 9.3 ± 1.5 |
| $^{56}$Co | 77.236 d | 6.0 ± 1.0 | 6.0 ± 0.5 | 6.2 ± 0.6 | 6.2 ± 0.8 | 6.6 ± 0.8 | 5.5 ± 0.7 | 5.6 ± 0.7 | 6.3 ± 1.2 |
| $^{46}$Sc | 83.787 d | 8.2 ± 1.0 | 8.2 ± 0.7 | 9.4 ± 0.8 | 8.3 ± 0.8 | 8.5 ± 0.8 | 8.7 ± 0.8 | 8.1 ± 0.8 | 10.1 ± 1.3 |
| $^{57}$Co | 271.8 d | 10.6 ± 1.4 | 10.5 ± 1.0 | 13.5 ± 1.4 | 12.4 ± 1.4 | 15.0 ± 1.6 | 9.5 ± 1.1 | 9.4 ± 1.2 | 13.9 ± 1.7 |
| $^{54}$Mn | 312.3 d | 88.3 ± 9.4 | 76.6 ± 7.7 | 92.5 ± 9.3 | 83.5 ± 8.5 | 83.8 ± 8.5 | 73.4 ± 7.5 | 70.2 ± 7.1 | 98.7 ± 10.2 |
| $^{22}$Na | 2.60 y | 87.6 ± 9.9 | 74.4 ± 5.7 | 91.0 ± 9.1 | 91.0 ± 7.2 | 84.6 ± 8.7 | 74.6 ± 5.9 | 77.7 ± 8.0 | 84.0 ± 6.9 |
| $^{60}$Co | 5.27 y | <5.3 | 0.6 ± 0.2 | 0.5 ± 0.2 | 1.3 ± 0.4 | 1.3 ± 0.3 | <2.9 | <2.1 | 1.4 ± 0.6 |
| $^{44}$Ti | 60 y | <8.9 | 1.3 ± 0.6 | <2.8 | <4.0 | <4.1 | <4.6 | <3.5 | <6.7 |
| $^{26}$Al | 7.17 × 10$^5$ y | 55.2 ± 6.7 | 42.0 ± 4.3 | 53.1 ± 5.4 | 50.5 ± 5.4 | 51.3 ± 5.4 | 47.4 ± 3.9 | 45.3 ± 4.8 | 51.5 ± 4.6 |
| $^{22}$Na/$^{26}$Al | | 1.6 ± 0.3 | 1.8 ± 0.2 | 1.7 ± 0.2 | 1.8 ± 0.2 | 1.7 ± 0.2 | 1.6 ± 0.2 | 1.7 ± 0.3 | 1.6 ± 0.2 |

## RESULTS

### Dark Flight Modeling

The results are summarized in Fig. 1, which shows the predicted impact locations and estimated final masses. As can be seen from this image, the multiple meteorite finds in the Ejby suburb appear to be the result of very late fragmentation at a height around 15 km, which is also supported by the very low amount of ablation on some of the surfaces of the Ejby fragments. Despite the fusion crust–free surfaces, none of the Ejby fragments fitted together. This suggests that not all of the fragments were found. In contrast to the Ejby fragments, the southernmost fragment found in the Glostrup area seems to originate from early fragmentation. The recovery location is consistent with our model calculation for fragment separating from the main mass at an altitude of 53 km. The recovered meteorites are all found systematically to the east of the predicted fall locations. We find that increasing the model wind speed by 10% will shift all calculated impact trajectories about 200 m to the east, which would make all finds, except the Herlev fragment, consistent with our dark-flight modeling. A few of the recovered meteorites were found in locations that are relatively far from the predicted fall areas. While one could imagine that very light fragments, such as the 18 g fragment, be carried farther east by gusts of the relatively strong winds that prevailed, the location of the main meteorite mass (6.5 kg) in Herlev is difficult to explain. Given the large mass of the Herlev fragment, it would have dominated the luminous trail. However, the distance between the find locations of the Herlev fragment and the location of the fragments found in the Ejby suburb is approximately 3.5 km. From the position of the German cameras, this distance corresponds to an angular offset of 0.65°. Such a large deviation in the luminous trail would have easily been noticed by these cameras (see fig. 5 in Bartoschewitz et al. 2017; Spurny et al. 2017). The Herlev fragment, therefore, must have followed the entire luminous path down to 18.3 km, and any deviations from other fragments must have been the result of aerodynamic effects during the darkflight phase.

There could be several reasons why the Herlev fragment experienced a lateral acceleration due to aerodynamic effects. It may have been possible that Herlev was a very nonspherical fragment that originated when the meteoroid split in two pieces at an altitude of 53 km, and thereafter experienced oriented fall. An alternative explanation may be that Herlev exhibited strong rotation leading to lateral acceleration (the Magnus effect). Unfortunately, the Herlev fragment shattered into many smaller pieces upon impact, which makes it difficult to test these hypotheses. However, what can be said is that none of the intact ablated surfaces of the Herlev fragment collection show clear signs of flow patterns typical for oriented ablation. Another explanation could be interaction of the Herlev fragment with the bow shock, as described in Passey and Melosh (1980). Without further speculation on the cause of the deviation of the Herlev fragments, we attempted to quantify the deviation with the simple assumption that some of the deceleration experienced by this fragment was not exactly opposite the direction of travel, but contained a lateral component. The advantage of this assumption is that the lateral deviation of the fragment is introduced very late in the trajectory because deceleration strongly peaks near the end of the luminous trajectory and at the beginning of dark flight. We find that we only need to invoke a lateral acceleration that is 11% of the deceleration in order to explain the Herlev impact location. This may also explain the eastward location of the 18 g fragment, which might have separated at a very late stage from the Herlev fragment.

**Mineralogy**

Based on the study of the thin sections prepared from the Herlev fragment, Ejby is unbrecciated. In our thin section, Ejby is highly recrystallized in most areas but also having parts in which a relict chondritic texture is preserved (Fig. 2a). Especially barred-olivine (BO) chondrules can be well recognized (Fig. 2b). Olivine is by far the most abundant phase and the grains are variable in size. Large plagioclase grains were also observed in Ejby indicating a high degree of recrystallization (Figs. 2c and 2d). Based on these features, this rock is transitional between a type 5 and type 6 chondrite. Ca-pyroxene has a much lower modal abundance compared to low-Ca pyroxene. As phosphates, both merrillite and Cl-apatite were found as accessory phases as well as chromites and ilmenites. A large chromite-rich aggregate is shown in Fig. 2e, which may result from metamorphic alteration of a former spinel-rich, Ca,Al-rich inclusion. During metamorphism, the spinels probably have gained abundant Fe and Cr exchanging Al and Mg by diffusion. A Cr-spinel-rich relic chondrule (Fig. 2f) certainly results from thermal alteration of an Al-chondrule and is very similar to the Cr-rich variety of Na,Al-rich chondrules described earlier (Bischoff and Keil 1983a, 1983b, 1984; Bischoff et al. 1989; Ebert and Bischoff 2016). Within this chondrule, olivine, apatite, merrillite, Cr-spinel, ilmenite, and kamacite are embedded in a fine-grained, Na-rich plagioclase-normative groundmass. These types of Alrich chondrules are also found in other type 5 ordinary chondrites (Bischoff and Keil 1983b, 1984). In places, high metal (kamacite, taenite, and tetrataenite) and sulfide abundances were found. Tetrataenite appears to be more abundant than taenite. Olivine shows undulatory extinction indicating that the rock is very weakly shocked (S2; Stöffler et al. 1991; Bischoff and Stöffler 1992).

The olivines are homogeneous in composition throughout the entire thin sections. The mean composition of 19 analyzed olivines is $Fa_{19.3\pm0.2}$ with a compositional range between 18.8 and 19.5 mol.% (Table 5). The low-Ca pyroxenes have a mean composition of $Fs_{16.9\pm0.2}$ (n = 21) varying between 16.4 and 17.2 mol.% in Fs-content; Ca-pyroxenes have a mean Fs-content of about 6 mole% (Table 5). Mean plagioclase has An- and Or-components of 11.5±0.9 and 6.0±1.6 mole%, respectively (n = 10; range An: 9.0–12.8 mole%). Chemical compositions of minor phases like apatite, merrillite, chromite, and ilmenite are given in Table 5.

Table 5. Chemical composition of main phases in the Ejby H5/6 ordinary chondrite.

| wt% | Olivine (n = 19) | Pyroxene (n = 21) | Ca-Pyroxene (n = 5) | Plagioclase (n = 10) | Apatite (n = 4) | Merrillite (n = 5) | Chromite (n = 4) | Ilmenite (n = 1) |
|---|---|---|---|---|---|---|---|---|
| $SiO_2$ | 39.1 | 56.3 | 54.5 | 65.2 | <0.08 | <0.03 | <0.04 | n.d. |
| $TiO_2$ | <0.03 | 0.17 | 0.47 | <0.05 | <0.03 | n.d. | 1.91 | 54.5 |
| $Al_2O_3$ | <0.01 | <0.15 | 0.69 | 21.5 | n.d. | n.d. | 6.4 | <0.04 |
| $Cr_2O_3$ | <0.02 | <0.10 | 0.95 | <0.02 | <0.05 | <0.01 | 57.5 | <0.10 |
| FeO | 18.2 | 11.4 | 3.61 | 0.35 | 0.36 | 0.38 | 28.7 | 38.3 |
| MnO | 0.48 | 0.51 | 0.22 | <0.01 | <0.02 | <0.04 | 0.88 | 1.92 |
| MgO | 42.7 | 31.0 | 17.0 | <0.02 | <0.02 | 3.57 | 2.97 | 4.42 |
| CaO | n.d. | 0.68 | 22.2 | 2.43 | 56.5 | 47.5 | <0.01 | n.d. |
| $Na_2O$ | <0.01 | <0.01 | 0.47 | 9.6 | 0.36 | 2.77 | n.d. | <0.05 |
| $K_2O$ | n.d. | <0.01 | <0.03 | 1.07 | <0.01 | <0.06 | n.d. | <0.03 |
| $P_2O_5$ | <0.02 | <0.01 | n.d. | <0.01 | 40.4 | 45.6 | n.d. | n.d. |
| Cl | n.d. | n.d. | n.d. | <0.01 | 4.42 | n.d. | n.d. | <0.01 |
| Total | 100.5 | 100.2 | 100.2 | 100.3 | 102.2 | 100.0 | 98.4 | 99.4 |
| Fa | 19.3 ± 0.2 | | | | | | | |
| Fs | | 16.9 ± 0.2 | 5.8 ± 0.4 | | | | | |
| Wo | | 1.3 ± 0.3 | 45.6 ± 0.5 | | | | | |
| An | | | | 11.5 ± 0.9 | | | | |
| Or | | | | 6.0 ± 1.6 | | | | |

Sample derived from fragment 3. All data in wt%.
n.d. = not detected; $n$ = number of analyses.

Kamacite has mean Ni- and Co-concentrations of 6.3 and 0.45 wt%, respectively (n = 9; Table 6), and tetrataenite has 49.3 ±2.0 wt% Ni (mean Co: ~0.05 wt%; n = 9). During analysis, only two other metals with 71 and 58 wt% Fe (taenite) were measured. The average of 10 troilite analyses is given in Table 6.

Table 6. Chemical composition of metals and troilite within the Ejby H5/6 ordinary chondrite.

| wt% | Kamacite (n = 9) | Taenite (n = 2) | Tetrataenite (n = 9) | Troilite (n = 10) |
|---|---|---|---|---|
| Fe | 94.8 ± 0.7 | 64.6 ± 6.9 | 51.1 ± 1.9 | 64.6 ± 0.3 |
| Ni | 6.3 ± 0.3 | 36.0 ± 6.8 | 49.3 ± 2.0 | <0.03 |
| Co | 0.45 ± 0.06 | 0.05 | 0.05 | <0.01 |
| S | n.d. | n.d. | n.d. | 36.3 ± 0.1 |
| Total | 101.6 | 100.6 | 100.5 | 100.9 |

All data in wt%. The sample was derived from fragment 3.
n.d. = not detected; $n$ = number of analyses.

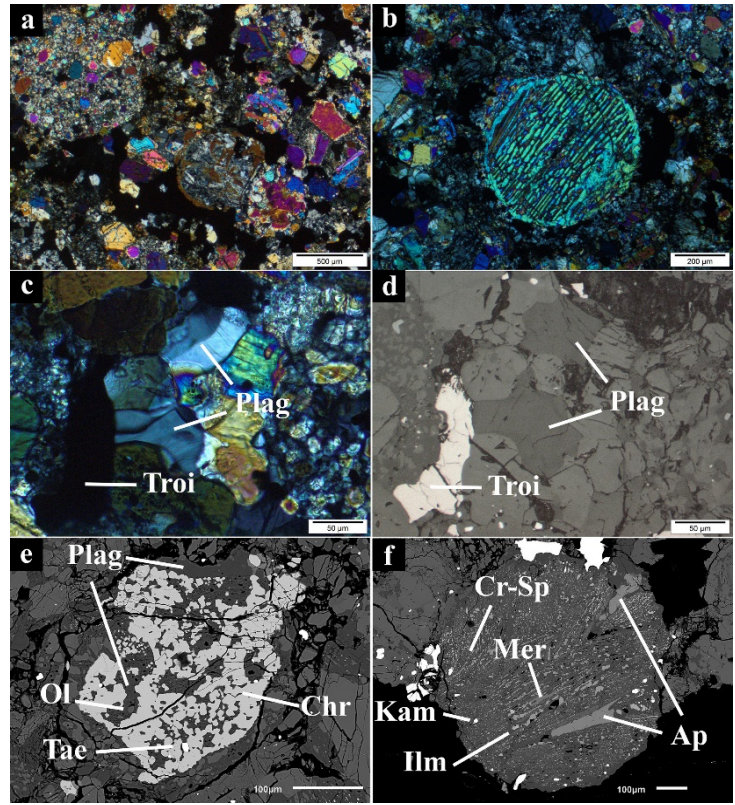

Figure 2: a) Photomicrograph of Ejby showing the recrystallized, but still chondritic texture, of the rock. b) relict barred-olivine chondrule (BO). c, d) Large plagioclases indicate the high metamorphic degree of the rock. e) A large chromite-rich aggregate, which may result from secondary recrystallization of a former spinel-rich CAI. f) Na,Al,Cr-rich relic chondrule. Images in polarized light, crossed nicols (a, b, c), reflected light (d), and in backscattered electrons (BSE; e, f). Plag = plagioclase; Troi = troilite; Ol = olivine; Tae = taenite; Chr = chromite; Kam = kamacite; Ilm = ilmenite; Met = merrillite; Ap = apatite; Cr-Sp = Cr-spinel.

**Oxygen Isotopes**

The oxygen isotope composition of the Ejby meteorite is given in Table 2 and plotted in Fig. 3 in relation to the H chondrite analyses of Clayton et al. (1991) and McDermott et al. (2016). The $\Delta^{17}O$ values for the McDermott et al. (2016) analyses were calculated using the same linearized method (Miller 2002) as the Ejby data, whereas the Clayton et al. (1991) $\Delta^{17}O$ values were calculated using the relation: $\Delta^{17}O = \delta^{17}O - 0.52 \times \delta^{18}O$. However, the two methods result in insignificant differences in $\Delta^{17}O$ values at the scale plotted in Fig. 3 and so the two datasets are essentially fully comparable. The Clayton et al. (1991) data show slightly more scatter than that of McDermott et al. (2016), but the compositional fields defined by both sets of analyses are very similar (Fig. 3). The mean of the two replicate analyses of Ejby plots within the H chondrite fields of both Clayton et al. (1991) and McDermott et al. (2016) (Fig. 3), thus providing further confirmation that Ejby is an H group chondrite. This result is consistent with its classification based on mineralogical criteria (see the Discussion section). The H chondrite data of McDermott et al. (2016) are plotted in Fig. 3 with respect to its petrologic type (H3–H6). However, the data show no clear evidence of systematic differences in oxygen isotope composition based on petrologic type, although the Ejby analysis does plot close to the relatively tight cluster defined by the four H6 samples analyzed by McDermott et al. (2016).

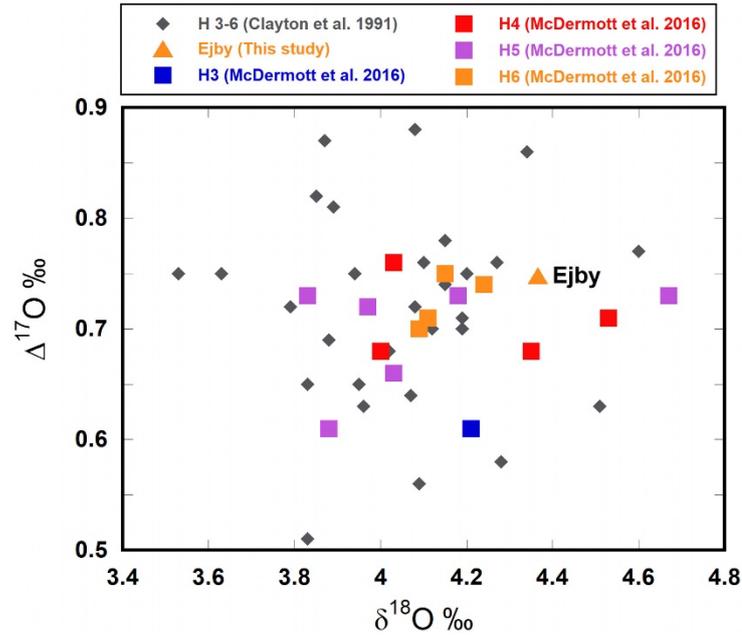

Figure 3: Oxygen isotope composition of Ejby shown in relation to the H chondrite data of Clayton et al. (1991) and McDermott et al. (2016).

**Titanium and Chromium Isotopes**

The Ti isotope data obtained for Ejby (n = 7; $\varepsilon^{46}Ti = -0.11 \pm 0.14$, $\varepsilon^{48}Ti = -0.06 \pm 0.09$, $\varepsilon^{50}Ti = -0.70 \pm 0.13$) agree within uncertainty with literature values for ordinary chondrites (Trinquier et al. 2009; Zhang et al. 2012; Williams 2015) (Fig. 4). The Cr isotope composition of Ejby is $0.194 \pm 0.025$ and $-0.423 \pm 0.060$ for $\varepsilon^{53}Cr$ and $\varepsilon^{54}Cr$, respectively. As for Ti isotopes, its Cr isotopic composition falls within the range defined by ordinary chondrites (Fig. 4) (Trinquier et al. 2007; Jenniskens et al. 2014; Bischoff et al. 2017).

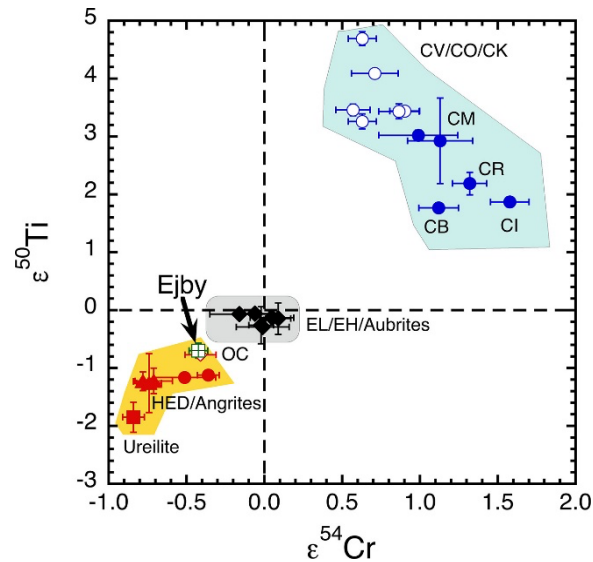

Figure 4: Titanium vs. Cr isotope data for various meteorite groups. Each group occupies a unique space in the diagram. Epsilon is given relative to the Earth. Titanium isotope data from this study and Williams (2015). Chromium isotope data from this study and Trinquier et al. (2007).

**Noble Gases**

Neon in Ejby is, as $^3$He and $^{38}$Ar, entirely cosmogenic (Tables 7 and 8), allowing us to use the measured $^{22}$Ne/$^{21}$Ne ratio of 1.1153 ± 0.0027 (Table 7) directly as shielding parameter. Both $^{22}$Ne/$^{21}$Ne ratios measured in Ejby small and large agree within uncertainty and we used the more precise ratio measured with the ion counter in the smaller fragment. Using the bulk chemistry for the major target elements (Table 3, average used where two values were given) and the production rates for He, Ne, and Ar isotopes determined for these elements in ordinary chondrites under various shielding conditions modeled by Leya and Masarik (2009), we can determine the CRE ages. These model predictions combined with the cosmogenic $^{22}$Ne/$^{21}$Ne ratio allow for a large range of shielding conditions for the Ejby meteoroid. With an assumed H chondrite density of 3.44 g cm3, the recovered mass of 8.98 kg requires the Ejby meteoroid to have been at least 8.5 cm in radius, in agreement with a minimum radius estimated by the examination of cosmogenic $^{26}$Al (see below), which can hence serve as a minimum shielding requirement.

The model predictions for the measured $^{22}$Ne/$^{21}$Ne ratio are consistent with preatmospheric radii of 20–120 cm and sample shielding depth of 4–8 cm. The ages averaged for these conditions are given in Table 9 and agree with each other (in contrast to those determined for 20–30 cm). This implies that there was no late loss of He relative to Ne or Ar during entry into the Earth's atmosphere or due to an orbit for any significant time close to the Sun. Also, the 38Ar-derived CRE ages do not deviate from each other or from the $^3$He- and $^{21}$Ne-derived CRE ages. Such deviations—beyond what is explainable by their uncertainties are often observed, mainly due to the large heterogeneity of Ca and the lack of consideration of the target element K in the model, but not here for Ejby. This suggests that these variations are most likely (1) not due to flawed production rates determined for the major target elements, and in particular Ca (Leya and Masarik 2009) and (2) that the major Ca-bearing minerals in Ejby are homogeneously distributed even at the 15 mg scale of the small sample. Alternatively, cosmogenic $^3$He could have experienced some (minor) loss compared to $^{21}$Ne, and the $^{38}$Ar-derived age could be lower than the $^{21}$Ne-derived one due to chemical heterogeneity, but we consider this not very likely.

From these results, we determine a CRE age of 88 ± 9 Ma based on noble gas systematics alone (Table 9). However, taking into account the results of dark-flight modeling, which determined the preatmospheric mass to ~120 kg and a maximum radius of ~20 cm (see above) and cosmogenic radionuclides (see above), we determined a similar set of CRE ages for a radius of 20–30 cm, which yielded smaller uncertainties for each of the albeit not concordant ages (Table 9). Our preferred CRE age for Ejby determined from these stricter conditions is 83 ± 11 Ma.

Correcting the measured $^4$He concentrations for cosmogenic He (we use [$^4$He/$^3$He]$_{cos}$ ~5.65 ± 0.45; Wieler 2002) yields a radiogenic $^4$He concentration of ~950 ± 60 cm$^3$ g$^{-1}$, a typical value for H chondrites. Combined with U and Th (Table 3), we obtain a U/ Th-$^{4}$He gas-retention age for Ejby of 3.1 ± 0.2 Ga.

We used K concentrations of 664 and 880 ppm, respectively (Table 3), to determine a nominal K-Ar age for Ejby. The radiogenic $^{40}$Ar detected in both aliquots was corrected for a very small (0–1% of the measured $^{40}$Ar) trapped $^{40}$Ar component (assuming [$^{40}$Ar/$^{36}$Ar]$_{tr}$ = 147.5 ± 147.5 covering both, primordially trapped Ar and air). The resulting nominal K-Ar age is in the range 4.5 ± 0.3 Ga, suggesting no major event affecting full Ar retentivity occurred after crystallization and metamorphism on the Ejby parent body.

Small amounts of trapped and cosmogenic Kr and Xe were detected (Tables 10 and 11). All Kr isotope data including $^{80}$Kr/$^{84}$Kr and $^{82}$Kr/$^{84}$Kr plot on a mixing line between the trapped Kr (Q or air, not precisely discernable due to the low count rates) and the expected cosmogenic endmember, implying that no neutroninduced Br-derived Kr is present, consistent with the comparably small preatmospheric radius and a low halogen abundance to be expected for type H5/6 chondrites. Likewise, an excess of n-derived $^{128}$Xe from I was not detected. However, the $^{129}$Xe/$^{132}$Xe ratios (Table 11, ~1.2) are high, relative to a pure mixture of trapped and cosmogenic Xe, proving that now extinct short-lived $^{129}$I was still present when the precursor material of Ejby closed for Xe loss, consistent with its very old nominal K-Ar age.

The small trapped Kr and Xe concentrations are typical for metamorphosed type 5/6 ordinary chondrites (cf. Alaerts et al. [1979] for LL chondrites). The $^{84}$Kr/$^{132}$Xe ratio of ~0.65, remarkably similar in both aliquots, suggests a small amount of phase Q gas (Busemann et al. 2000) remaining after metamorphism, while trapped He to Ar are essentially lost. The $^{84}$Kr/$^{132}$Xe ratio is much lower than what would be expected for strongly weathered gas-poor metamorphosed ordinary chondrites (which should carry much higher $^{84}$Kr/$^{132}$Xe values due to the presence of air). This may reflect the fast recovery of the stone which minimized the impact of contamination.

We can correct the measured Kr and Xe for trapped gases (we assumed a mixture of Q and air) and use the resulting cosmogenic $^{83}$Kr and $^{126}$Xe to determine the CRE ages. Using the shielding-dependent production rates given for H chondrites by Eugster (1988), we obtain CRE ages of 86 and 94 Ma for $^{126}$Xe and 84 and 78 Ma for $^{83}$Kr for the small and large samples, respectively. This is strikingly similar to the preferred age of 83 ± 11 Ma determined for the light cosmogenic noble gases.

Table 7. Helium and Ne concentrations (in $10^{-8}$ cm$^3$ g$^{-1}$ STP) and isotopic ratios in two specimens "small" (15.41 ± 0.01 mg) and "large" (117.82 ± 0.01 mg) of the Ejby meteorite, fragment 3.

| | $^3$He$_{cos}$ | $^4$He | $^3$He/$^4$He × 10,000 | $^{20}$Ne | $^{20}$Ne/$^{22}$Ne | $^{21}$Ne/$^{22}$Ne | $^{21}$Ne$_{cos}$ | $^4$He$_{rad}$ |
|---|---|---|---|---|---|---|---|---|
| Small | 128.2 ± 2.2 | 1635 ± 17 | 784 ± 16 | 25.89 ± 0.11 | 0.8087 ± 0.0026 | 0.8966 ± 0.0022 | 28.70 ± 0.10 | 910 ± 61 |
| Large | 129.7 ± 2.1 | 1722 ± 18 | 753 ± 15 | 26.43 ± 0.15 | 0.82 ± 0.03 | 0.90 ± 0.04 | 29.1 ± 0.6 | 990 ± 62 |

Cos = cosmogenic; rad = radiogenic.

Table 8. Argon concentrations (in $10^{-8}$ cm$^3$ g$^{-1}$) and isotopic ratios in two specimens "small" (15.41 ± 0.01 mg) and "large" (117.82 ± 0.01 mg) of the Ejby meteorite, fragment 3.

| | $^{36}$Ar | $^{36}$Ar/$^{38}$Ar | $^{40}$Ar/$^{36}$Ar | $^{38}$Ar$_{cos}$ | $^{40}$Ar$_{rad}$ |
|---|---|---|---|---|---|
| Small | 2.85 ± 0.08 | 0.745 ± 0.019 | 2050 ± 60 | 3.750 ± 0.016 | 5790 ± 230 |
| Large | 2.40 ± 0.05 | 0.624 ± 0.026 | 2520 ± 50 | 3.878 ± 0.017 | 6080 ± 180 |

cos = cosmogenic; rad = radiogenic.

Table 9. Model production rates $P_x$ (Leya and Masarik 2009) for $^3$He, $^{21}$Ne, and $^{38}$Ar based on the shielding parameter ($^{22}$Ne/$^{21}$Ne)$_{cos}$ = 1.1153 ± 0.0027 measured in Ejby (see text) and the chemistry given in Table 3. Two production rate and CRE age ($T_x$) sets are given (1) averaged for a preatmospheric size of 20–120 cm radius, based on noble gas constraints alone and (2) based on the external estimate of a preatmospheric radius of 20–30 cm by dark flight modeling and radionuclides, which we consider our preferred CRE age for Ejby.

| | $P_3$ 20–120 cm | $P_{21}$ | $P_{38}$ | $P_3$ 20–30 cm | $P_{21}$ | $P_{38}$ | $T_{avrg}$ 20–120 cm | $T_{avrg}$ 20–30 cm |
|---|---|---|---|---|---|---|---|---|
| | 1.59 ± 0.28 | 0.29 ± 0.04 | 0.046 ± 0.008 | 1.707 ± 0.009 | 0.297 ± 0.010 | 0.0493 ± 0.0004 | | |

| | $T_3$ 20–120 cm | $T_{21}$ | $T_{38}$ | $T_3$ 20–30 cm | $T_{21}$ | $T_{38}$ | | |
|---|---|---|---|---|---|---|---|---|
| Small | 81 ± 14 | 99 ± 15 | 81 ± 14 | 75 ± 1 | 97 ± 3 | 76 ± 1 | | |
| Large | 81 ± 15 | 100 ± 15 | 84 ± 14 | 76 ± 1 | 98 ± 4 | 79 ± 1 | 88 ± 9 | 83 ± 11 |

Production rates $P_x$ in $10^{-8}$ cm$^3$/(g × Ma), exposure ages $T_x$ in Ma.

Table 10. Krypton concentration and isotopic ratios in two specimens "small" (15.41 ± 0.01 mg) and "large" (117.82 ± 0.01 mg) of the Ejby meteorite, fragment 3.

| | $^{84}$Kr $10^{-10}$ cm$^3$ g$^{-1}$ | $^{78}$Kr/$^{84}$Kr $^{84}$Kr = 100 | $^{80}$Kr/$^{84}$Kr | $^{82}$Kr/$^{84}$Kr | $^{83}$Kr/$^{84}$Kr | $^{86}$Kr/$^{84}$Kr |
|---|---|---|---|---|---|---|
| Small | 0.83 ± 0.04 | 2.30 ± 0.13 | 10.3 ± 0.5 | 28.0 ± 1.3 | 30.9 ± 1.3 | 28.4 ± 1.4 |
| Large | 0.706 ± 0.020 | 2.607 ± 0.029 | 11.86 ± 0.13 | 29.5 ± 0.4 | 31.7 ± 0.5 | 27.85 ± 0.27 |

Table 11. Xenon concentration and isotopic ratios in two specimens "small" (15.41 ± 0.01 mg) and "large" (117.82 ± 0.01 mg) of the Ejby meteorite, fragment 3.

| | $^{132}$Xe $10^{-10}$ cm$^3$ g$^{-1}$ | $^{124}$Xe/$^{132}$Xe $^{132}$Xe = 100 | $^{126}$Xe/$^{132}$Xe | $^{128}$Xe/$^{132}$Xe | $^{129}$Xe/$^{132}$Xe | $^{130}$Xe/$^{132}$Xe | $^{131}$Xe/$^{132}$Xe | $^{134}$Xe/$^{132}$Xe | $^{136}$Xe/$^{132}$Xe |
|---|---|---|---|---|---|---|---|---|---|
| Small | 1.28 ± 0.06 | 0.79 ± 0.04 | 0.87 ± 0.05 | 8.9 ± 0.5 | 115 ± 6 | 16.7 ± 1.0 | 83 ± 4 | 37.6 ± 2.0 | 32.0 ± 1.7 |
| Large | 1.087 ± 0.027 | 0.762 ± 0.021 | 1.002 ± 0.013 | 9.13 ± 0.10 | 122.5 ± 1.3 | 16.73 ± 0.19 | 83.3 ± 1.0 | 37.9 ± 0.5 | 32.3 ± 0.4 |

Table 12. Concentration of primordial radionuclides (ng g$^{-1}$ for U and Th chains and for $K_{nat}$) in eight specimens of the Ejby meteorite measured by nondestructive gamma-ray spectroscopy. Errors include a 1σ uncertainty of 10% in the detector efficiency calibration.

| Fragment | #2 | #3a | #3b | #3c | #4 | #5 | #11 |
|---|---|---|---|---|---|---|---|
| Mass (g) | 314.25 | 200.18 | 85.58 | 82.57 | 64.59 | 58.5 | 18.1 |
| $^{238}$U | 10 ± 2 | 9 ± 2 | 13 ± 3 | 20 ± 2 | 13 ± 1 | 9 ± 1 | 13 ± 2 |
| $^{228}$Th | 35 ± 5 | 35 ± 6 | 40 ± 7 | 32 ± 3 | 45 ± 4 | 28 ± 3 | 28 ± 4 |
| $^{40}$K | 690 ± 100 | 680 ± 100 | 720 ± 100 | 650 ± 70 | 820 ± 80 | 720 ± 70 | 790 ± 80 |

**Cosmogenic Radionuclides**

Data on the short-lived cosmogenic radionuclides are given in Tables 4 and 12. The activities for the very short-lived radionuclides are low (see Table 4) and the naturally occurring radionuclides (Table 12) are in agreement with the average concentrations in ordinary H chondrites (Wasson and Kallemeyn 1988).

**Soluble Organics Chemistry**

Data on the soluble organics chemistry are presented in Fig. 5. These data show a high amount of endogene thermostable soluble organic matter. The highest signals show the typical organic meteoritic signature of longchain fatty acids and sulfonated alkanes (Schmitt- Kopplin et al. 2010; Haack et al. 2012; Bischoff et al. 2017). Ejby also contains organomagnesium compounds that are formed preferentially in high metamorphic meteorite samples (Ruf et al. 2017) (Fig. 5).

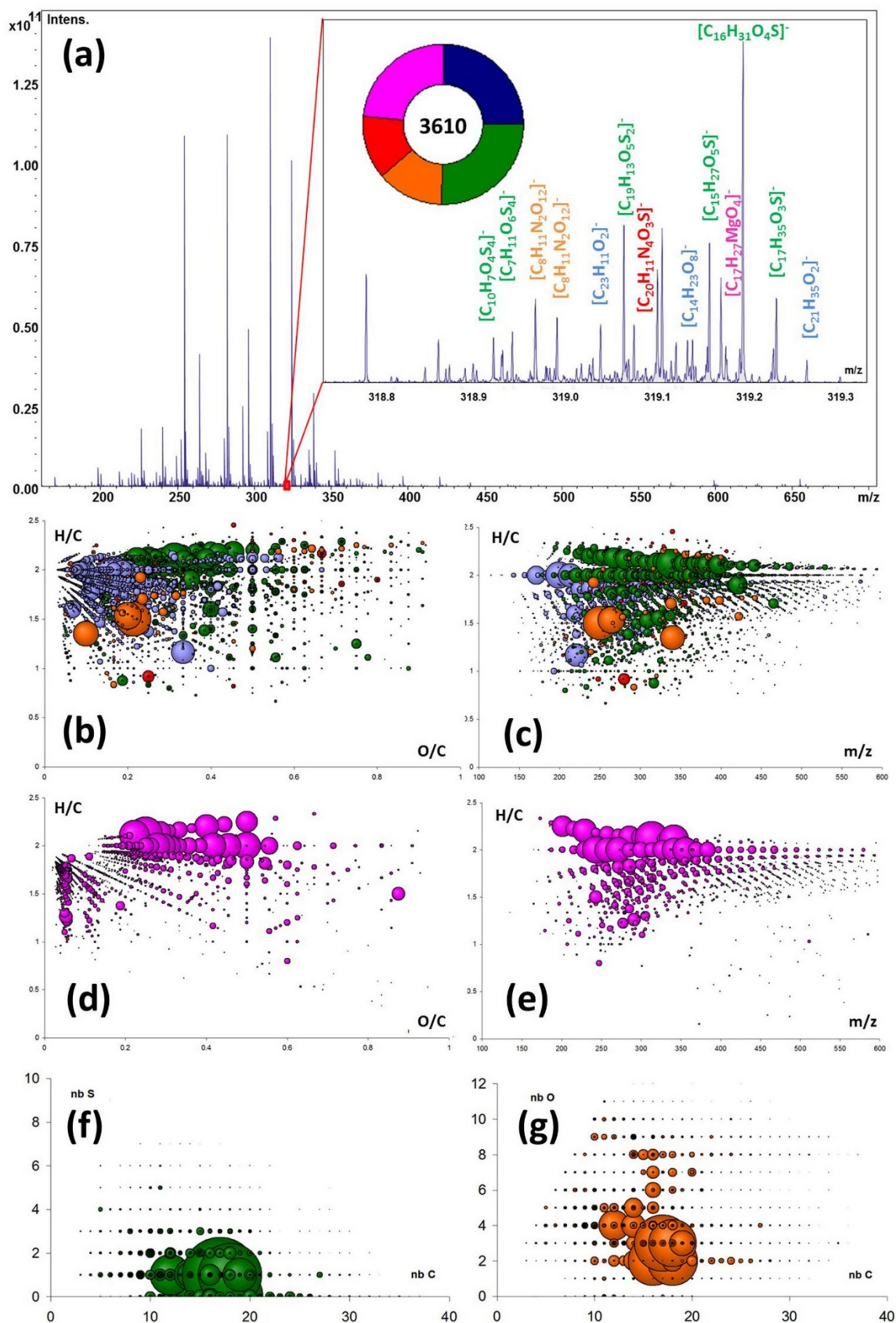

Figure 5: Methanol soluble organic compounds show (a) a very dense mass spectra with 3610 annotations in the CHNOS space, (b, c) the van Krevelen diagrams show the prevalence of CHO and CHOS compounds with homologous chemical structures up to higher mass range (600 amu.), (d, e) the organomagnesium compounds have a parallel structural profile to the CHO with additional low oxygenated and unsaturated CHOMg-series, (f) the CHOS and (g) the CHNO show high amount of sulfur and nitrogen in the formula and nitrogen is present in molecules of higher number of carbons.

## DISCUSSION

### Classification of Ejby

Ejby is an unbrecciated chondritic rock on the thin section scale. The rock is well recrystallized, but the chondritic texture is still visible in thin section (Figs. 2a and 2b). The recrystallized texture already points to a high metamorphic grade of the rock. The large grain size of plagioclase (partly >50 lm; Fig. 2) and the homogeneous compositions of olivine and pyroxenes (Table 5) also support this. These facts clearly indicate that the Ejby ordinary chondrite is intermediate between type 5 and type 6. The oxygen isotope composition of Ejby (Fig. 3) is consistent with it being an H chondrite, in keeping with the mean olivine and low-Ca pyroxene compositions of $Fa_{19.3}$ and $Fs_{16.9}$, respectively. Titanium isotopes are also consistent with the classification as an ordinary chondrite (Fig. 4) and trapped Kr and Xe concentrations are comparable to concentrations in type 5 or 6 (LL) ordinary chondrites (Alaerts et al. 1979). As indicated by the undulatory extinction seen in olivine and plagioclase, Ejby is very weakly shocked (S2), which is also consistent with the abundant presence of radiogenic 4He and 40Ar and so excludes any significant resetting of these gas-retention systems within the last 3 Ga. Thus, Ejby should be classified as a H5/6, S2 ordinary chondrite (Bouvier et al. 2017).

### A Rather Small Ejby Meteoroid

Negligible activity of $^{60}$Co (<1 dpm kg _1) and the lack of n-induced excesses in $^{80,82}$Kr and $^{128}$Xe suggest that, in spite of being a multiple fall, the preatmospheric size of the Ejby meteoroid was rather small and no significant production of secondary thermal neutrons took place within the meteoroid during its recent CRE in space. The measured $^{26}$Al activity is consistent with that expected for a moderatesize H chondrite (Bhandari et al. 1989; Bonino et al. 2001; Leya and Masarik 2009).

When we compare the radionuclide concentrations with cosmic ray production estimations for $^{26}$Al (Leya and Masarik 2009) (Fig. 6), $^{60}$Co (Eberhardt et al. 1963; Spergel et al. 1986), $^{54}$Mn (Kohman and Bender 1967), and $^{22}$Na (Bhandari et al. 1993), the best agreement is obtained (in the sequence of the given isotopes) for radii of 10–20 cm, <20 cm, 8–12 cm, and 10–20 cm. Combining all results of these radionuclides, we infer a roughly spherical meteoroid with a ~20 cm radius. The $^{22}$Na/$^{26}$Al ratio of (1.7 ± 0.1) is close to the average value for H chondrites (Bhandari et al. 1989). The cosmogenic $^{22}$Ne/$^{21}$Ne ratio determined in Ejby (Table 7) is consistent with a preatmospheric minimum radius of 20 cm determined with the radionucleides, although larger radii (up to 120 cm) are in agreement with this measured shielding parameter, too. A radius of 20 cm is also consistent with the fireball observation which suggests a radius of 20–30 cm (Spurny et al. 2017) and the results of the dark flight modelling (above) which suggest a radius of approximately 20 cm. The activities of the short-lived radioisotopes, with half-life less than the orbital period, represent the production integrated over the last segment of the orbit. The fall of the Ejby H6 chondrite occurred during the current solar cycle 24 maximum, as indicated by the neutron monitor data (Bartol Neutron Monitors 2016). The cosmic ray flux was low in the 6 months before the fall. Hence, the activities for the very short-lived radionuclides are expected to be low (see Table 4), as earlier reported (Bischoff et al. [2011] and references cited therein).

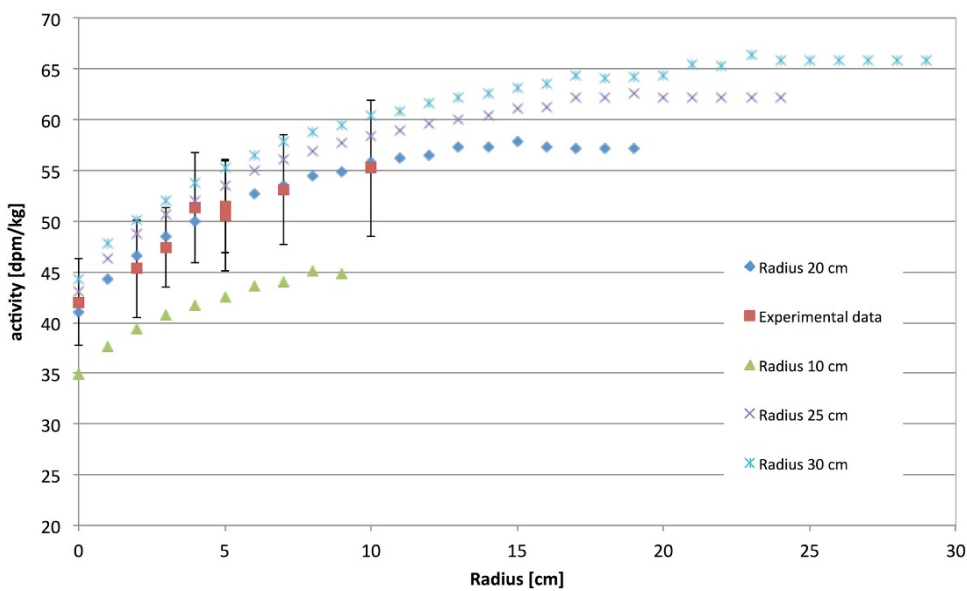

Figure 6: Comparison of modeling result from Leya and Masarik (2009) with experimental results from gamma-ray spectrometry for $^{26}$Al. The measured $^{26}$Al activities can be fitted to the modeling results for a spherical body with a diameter of 20 cm. (Color figure can be viewed at wileyonlinelibrary.com.)

The noble gas CRE age of 83 ± 11 Ma determined for Ejby is the highest age determined for an H chondrite so far (previous max. 77 Ma) and the second highest age for an ordinary chondrite. The only ordinary chondrite with a higher exposure age (95 Ma) is Y-74035 (Takaoka et al. 1981). CRE ages of other ordinary chondrites cover a range of 1 to ~80 Ma (Marti and Graf 1992; Graf and Marti 1995; Herzog and Caffee 2014). This CRE age for Ejby may represent roughly the possible maximum exposure to space erosion and destructive breakup events of stony meteoroids that are sufficiently small to still record GCR effects in its interior (e.g., Eugster et al. 2006).

**Significance of the Soluble Organics Chemistry**

We show herein that H chondrites of even highly metamorphic grade have a large amount of endogene thermostable soluble organic matter. The solvent used is the protic/polar methanol and the extracted compounds correspond to polar compounds ionizable in electrospray ionization mode prior to mass spectrometry. Exact mass spectrometry showed a very dense signature in each nominal mass in a high-intensity dynamic range and signals in the negative mass defect corresponding to high number of sulfur and oxygen (Fig. 5a). The highest signals show the typical organic meteoritic signature of long-chain fatty acids and sulfonated alkanes. The 3610 CHNOSMg elementary composition calculated from exact mass analysis could be visualized in their relative abundances in van Krevelen diagrams and showed a richness in aliphatic, sulfurized, and oxygenated compounds (Figs. 5b–e).

As demonstrated earlier (Ruf et al. 2017), organomagnesium compounds can be formed preferentially in highly metamorphic meteorites (see also Bartoschewitz et al. [2017] and Bischoff et al. [2017] concerning the falls of Braunschweig and Stubenberg for comparison); here as well complete signatures in CHOMg compounds formed during metamorphism of the Ejby parent body. These show highly aliphatic oxygenated as well as aromatic low oxygenated structures with low H/C and low O/C values.

**Concluding Remarks**

Together with the meteorite falls of Braunschweig (L6; Bartoschewitz et al. 2017), _Zd'_ar nad S_azavou (L3; Spurny et al. 2016), Stubenberg (LL6; Bischoff et al. 2017), Hradec Kr_alov_e (LL5; Bouvier et al. 2017), and Broek in Waterland (Bouvier et al. 2017), Ejby (H5/6) is the sixth recovered meteorite fall within less than 3 years impacting Earth only ~800 km apart. In comparison, during the last decades, less than 10 meteorite falls/year are recovered and reported globally. Thus, this concentration of recorded falls in Central Europe is remarkable.

In Denmark, the statistics are even more remarkable. In the previous two centuries, only one fall was recovered per century in Denmark; however, in this century, two falls (Maribo [CM2]; Haack et al. 2012) and Ejby have been recovered. Additionally, another meteorite (Dueodde, H5) where "the meteorite fall must have occurred within the last 2–3 years" was recently located under a leaking garage roof.

The significantly improved recovery statistics are likely mainly due to the growing number of all-sky camera networks and surveillance cameras in Europe. All of the above mentioned falls were recorded on either all-sky cameras (_Žd'ár nad Sázavou, Stubenberg, Hradec Králové) or on still images and/or video intended for other purposes. Increased public awareness of meteorites also played a role in the recovery of Ejby and Dueodde.

The Ejby fall is also unusual in the sense that the recovered mass is very high compared to the estimated mass of the meteoroid. Almost 9 kg was recovered of the original ≈110 kg, which is almost 8%. The high percentage is in part due to the low entry velocity and the fact that it fell in a major city, but public awareness also played a role in the recovery of such a high fraction of the original mass. In contrast, the recovered mass of Park Forest L5 ordinary chondrite, which fell in the suburbs of Chicago in 2003, was only between 0.5 and 3% of the original mass (Simon et al. 2004).

*Acknowledgments*: We thank the reviewers Alan Hildebrand and Greg Herzog and the Associate Editor Ed Scott for fruitful comments, suggestions, and support. We thank U. Heitmann (Münster) for thin section preparation. Knud-Jacob Simonsen is thanked for providing wind data from the Danish Meteorological Institute. This work has partly been carried out within the framework of the NCCR PlanetS supported by the Swiss National Science Foundation, and partly supported by the German Research Foundation (DFG) within the SFB-TRR 170 "Late Accretion onto Terrestrial Planets" (subproject B5). This is TRR 170 publication no. 58.

**References:**

Alaerts L., Lewis R. S., and Anders E. 1979. Isotopic anomalies of noble gases in meteorites and their origins—III. LL-chondrites. Geochimica et Cosmochimica Acta 43:1399–1415.

Arpesella C. 1996. A low background counting facility at Laboratori Nazionali del Gran Sasso. Applied Radiation and Isotopes 47:991–996.

Barrat J.-A., Zander B., Moynier F., Bollinger C., Liorzou C., and Bayron G. 2012. Geochemistry of CI chondrites: Major and trace elements, and Cu and Zn isotopes. Geochimica et Cosmochimica Acta 83:79–92.

Barrat J.-A., Gillet P., Dauphas N., Bollinger C., Etoubleau J., Bischoff A., and Yamaguchi A. 2016. Evidence from Tm anomalies for non-CI refractory lithophile element proportions in terrestrial planets and achondrites. Geochimica et Cosmochimica Acta 176:1–17.

Bartol Neutron Monitors. 2016. http://neutronm.bartol.udel.ed u/

Bartoschewitz R., Appel P., Barrat J.-A., Bischoff A., Caffee M. W., Franchi I. A., Gabelica Z., Greenwood R. C., Harir M., Harries D., Hochleitner R., Hopp J., Laubenstein M., Mader B., Marques R., Morlok A., Nolze G., Prudêncio M. I., Rochette P., Ruf A., Schmitt- Kopplin P., Seemann E., Szurgot M., Tagle R., Wach R. A., Welten K. C., Weyrauch M., and Wimmer K. 2017. The Braunschweig meteorite—A recent L6 chondrite fall in Germany. Chemie der Erde Geochemistry 77:207–224.

Bhandari N., Bonino G., Callegari E., Cini Castagnoli G., Mathew K. J., Padia J. T., and Queirazza G. 1989. The Torino H6 meteorite shower. Meteoritics 24:29–34. Bhandari N., Mathew K. J., Rao M. N., Herpers U., Bremer K., Vogt S., Wölfli W., Hofmann H. J., Michel R., Bodemann R., and Lange H.-J. 1993. Depth and size dependence of cosmogenic nuclide production rates in stony meteoroids. Geochimica et Cosmochimica Acta 57:2361–2375.

Bischoff A. and Keil K. 1983a. Ca-Al-rich chondrules and inclusions in ordinary chondrites. Nature 303:588–592.

Bischoff A. and Keil K. 1983b. Catalog of Al-rich chondrules, inclusions and fragments in ordinary chondrites. Special publication No. 22. Albuquerque, New Mexico: University of New Mexico, Institute of Meteoritics. pp. 1–33.

Bischoff A. and Keil K. 1984. Al-rich objects in ordinary chondrites: Related origin of carbonaceous and ordinary chondrites and their constituents. Geochimica et Cosmochimica Acta 48:693–709.
Bischoff A. and Stöffler D. 1992. Shock metamorphism as a fundamental process in the evolution of planetary bodies: Information from meteorites. European Journal of Mineralogy 4:707–755.

Bischoff A., Palme H., and Spettel B. 1989. Al-rich chondrules from the Ybbsitz H4 chondrite: Evidence for formation by collision and splashing. Earth and Planetary Science Letters 93:170–180.

Bischoff A., Jersek M., Grau T., Mirtic B., Ott U., Ku_cera J., Horstmann M., Laubenstein M., Herrmann S., Randa Z., Weber M., and Heusser G. 2011. Jesenice—A new meteorite fall from Slovenia. Meteoritics & Planetary Science 46:793–804.

Bischoff A., Barrat J.-A., Bauer K., Burkhardt C., Busemann H., Ebert S., Gonsior M., Hakenm€uller J., Haloda J., Harries D., Heinlein D., Hiesinger H., Hochleitner R., Hoffmann V., Kaliwoda M., Laubenstein M., Maden C., Meier M. M. M., Morlok A., Pack A., Ruf A., Schmitt- Kopplin P., Schönbächler M., Steele R. C. J., Spurny P., and Wimmer K. 2017. The Stubenberg meteorite—An LL6 chondrite fragmental breccia recovered soon after precise prediction of the strewn field. Meteoritics & Planetary Science 52:1683–1703.

Bonino G., Bhandari N., Murty S. V. S., Mahajan R. R., Suthar K. M., Shukla A. D., Shukla P. N., Cini Castagnoli G., and Taricco C. 2001. Solar and galactic cosmic ray records of the Fermo (H) chondrite regolith breccia. Meteoritics & Planetary Science 36:831–839.

Bouvier A., Gattacceca J., Grossman J., and Metzler K. 2017. The Meteorite Bulletin, No. 105. Meteoritics & Planetary
Science, 52: 2411. https://doi.org/10.1111/maps.12944

Busemann H., Baur H., and Wieler R. 2000. Primordial noble gases in "phase Q" in carbonaceous and ordinary chondrites studied by closed-system stepped etching. Meteoritics & Planetary Science 35:949–973.

Carter R. T., Jandir P. S., and Kress M. E. 2009. Estimating the drag coefficients of meteorites for all mach number regimes (abstract #2059). 40th Lunar and Planetary Science Conference. CD-ROM.

Clayton R. N., Mayeda T. K., Goswami J. N., and Olsen E. J. 1991. Oxygen isotope studies of ordinary chondrites. Geochimica et Cosmochimica Acta 55:2317–2337.

Eberhardt P., Geiss J., and Lutz H. 1963. Neutrons in meteorites. In Earth science and meteoritics, edited by Geiss J. And Goldberg E. D. Amsterdam: North Holland Publishing Company. pp. 143–168.

Ebert S. and Bischoff A. 2016. Genetic relationship between Na-rich chondrules and Ca, Al-rich inclusions? Formation of Na-rich chondrules by melting of refractory and volatile precursors in the solar nebula. Geochimica et Cosmochimica Acta 177:182–204.

Eugster O. 1988. Cosmic-ray production rates for $^3$He, $^{21}$Ne, $^{38}$Ar, $^{83}$Kr, and $^{126}$Xe in chondrites based on $^{81}$Kr-Kr exposure ages. Geochimica et Cosmochimica Acta 52:1649–1662.

Eugster O., Herzog G. F., Marti K., and Caffee M. E. 2006. Irradiation records, cosmic-ray exposure ages, and transfer times of meteorites. In Meteorites and the early solar system II, edited by Lauretta D. S., Leshin L. A., and McSween H. Y. Tucson, Arizona: University of Arizona Press. pp. 829–851.

Graf T. and Marti K. 1995. Collisional records in H chondrites. Journal of Geophysical Research 100:21,247–21,263.

Greenwood R. C., Burbine T. H., Miller M. F., and Franchi I. A. 2017. Melting and differentiation of early-formed asteroids: The perspective from high precision oxygen isotope studies. Chemie der Erde 7:1–43.

Gritsevich M. I. 2009. Determination of parameters of meteor bodies based on flight observational data. Advances in Space Research 44:323–334.

Haack H., Grau T., Bischoff A., Horstmann M., Wasson J. T., Sørensen A., Laubenstein M., Ott U., Palme H., Gellissen M., Greenwood R. C., Pearson V. P., Franchi I. A., Gabelica Z., and Schmitt-Kopplin P. 2012. Maribo—A new CM fall from Denmark. Meteoritics & Planetary Science 47:30–50.

Herzog G. F. and Caffee M. 2014. Cosmic-ray exposure ages of meteorites. In Planets, asteroids, comets and the solar system, 2nd ed., edited by Davis A. M. Treatise on Geochemistry, vol. 2. Oxford: Elsevier. pp. 419–453.

Jenniskens P., Rubin A. E., Yin Q. Z., Sears D. W., Sandford S. A., Zolensky M. E., Krot A. N., Blair L., Kane D., Utas J., and Verish R. 2014. Fall, recovery, and characterization of the Novato L6 chondrite breccia. Meteoritics & Planetary Science 49:1388–1425.

Kohman T. P. and Bender M. L. 1967. Nuclide production by cosmic rays in meteorites and on the Moon. In Highenergy nuclear reactions in astrophysics—A collection of articles, edited by Shen B. S. P. New York: W. A. Benjamin. pp. 169–245.

Laubenstein M. 2017. Screening of materials with high purity germanium detectors at the Laboratori Nazionali del Gran Sasso. International Journal of Modern Physics A 32:1743002.

Leya I. and Masarik J. 2009. Cosmogenic nuclides in stony meteorites revisited. Meteoritics & Planetary Science 44:1061–1086.

Marti K. and Graf T. 1992. Cosmic-ray exposure history of ordinary chondrites. Annual Review of Earth and Planetary Science 20:221–243.

McDermott K. H., Greenwood R. C., Scott E. R. D., Franchi I. A., and Anand M. 2016. Oxygen isotope and petrological study of silicate inclusions in IIE iron meteorites and their relationship with H chondrites. Geochimica et Cosmochimica Acta 173:97–113.

Miller M. F. 2002. Isotopic fractionation and the quantification of $^{17}$O anomalies in the oxygen three isotopes system: An appraisal and geochemical significance. Geochimica et Cosmochimica Acta 66:1881–1889.

Miller M. F., Franchi I. A., Sexton A. S., and Pillinger C. T. 1999. High precision $\delta^{17}$O isotope measurements of oxygen from silicates and other oxides: Method and applications. Rapid Communications in Mass Spectrometry 13:1211–1217.

Passey Q. R. and Melosh H. J. 1980. Effects of atmospheric breakup on crater field formation. Icarus 42:211–233.

Pecina P. and Ceplecha Z. 1983. New aspects in single-body meteor physics: Institutes of Czechoslovakia. Bulletin 34:102–121.

Riebe M. E. I., Welten K. C., Meier M. M. M., Wieler R., Bart M. I. F., Ward D., Laubenstein M., Bischoff A., Caffee M. W., Nishiizumi K., and Busemann H. 2017. Cosmic-ray exposure ages of six chondritic Almahata Sitta fragments. Meteoritics & Planetary Science 52:2353–2374.

Ruf A., Kanawati B., Hertkorn N., Yin Q.-Z., Moritz F., Harir M., Lucio M., Michalke B., Wimpenny J., Shilobreeva S., Bronsky B., Saraykin V., Gabelica Z., Gougeon R. D., Quirico E., Ralew S., Jakubowski T., Haack H., Gonsior M., Jenniskens P., Hinman N. W., and Schmitt-Kopplin P. 2017. Previously unknown class of metalorganic compounds revealed in meteorites. Proceedings of the National Academy of Science 114:2819–2824.

Schmitt-Kopplin P., Gabelica Z., Gougeon R. D., Fekete A., Kanawati B., Harir M., Gebefuegi I., Eckel G., and Hertkorn N. 2010. High molecular diversity of extraterrestrial organic matter in Murchison meteorite revealed 40 years after its fall. Proceedings of the National Academy of Science 107:2763–2768.

Simon S. B., Grossman L., Clayton R. N., Mayeda T. K., Schwade J. R., Sipiera P. P., Wacker J. F., and Wadhwa M. 2004. The fall, recovery and classification of the Park Forest meteorite. Meteoritics & Planetary Science 39:621–634.

Spergel M. S., Reedy R. C., Lazareth O. W., Levy P. W., and Slatest L. A. 1986. Cosmogenic neutron-capture-produced nuclides in stony meteorites. 16th Proceedings of the Lunar & Planetary Science Conference. Journal of Geophysical Research (Suppl. 91): D483–D494.

Spurny P., Borovi_cka J., Haloda J., Shrbeny L., and Heinlein D.. 2016. Two very precisely instrumentally documented meteorite falls: Zdar nad Sazavou and Stubenber Prediction and reality (abstract #6221). 79th Annual Meeting of theMeteoritical Society, LPI Contribution No. 1921.

Spurny P., Borovi_cka J., Baumgarten G., Haack H., Heinlein D., and Sørensen A. N. 2017. Atmospheric trajectory and heliocentric orbit of the Ejby meteorite fall in Denmark on February 6, 2016. Planetary and Space Science 143:192–198.

Stöffler D., Keil D., and Scott E. R. D. 1991. Shock metamorphism of ordinary chondrites. Geochimica et Cosmochimica Acta 55:3845–3867.

Takaoka N., Saito K., Ohba Y., and Nagao K. 1981. Rare gas studies of twenty-four Antarctic chondrites. In Proceedings of the Sixth Symposium on Antarctic Meteorites, edited by Nagata T. Memoirs of the National Institute of Polar Research, Special Issue No. 20. Tokyo: National Institute of Polar Research. pp. 264–275.

Trinquier A., Birck J.-L., and All_egre C. J. 2007. Widespread 54Cr heterogeneity in the inner solar system. The Astrophysical Journal 655:1179–1185.

Trinquier A., Elliott T., Ulfbeck D., Coath C., Krot A. N., and Bizzarro M. 2009. Origin of nucleosynthetic isotope heterogeneity in the solar protoplanetary disk. Science 324:374–376.

Wasson J. T. and Kallemeyn G. W. 1988. Compositions of chondrites. Philosophical Transactions of the Royal Society of London. Series A, Mathematical and Physical Sciences 325:535–544.

Wieler R. 2002. Cosmic-ray-produced noble gases in meteorites. Reviews in Mineralogy and Geochemistry 47:125–170.

Williams N. H. 2015. The origin of titanium isotopic anomalies within solar system material. Ph.D. thesis, The University of Manchester.

Zhang J., Dauphas N., Davis A. M., Leya I., and Fedkin A. 2012. The proto-Earth as a significant source of lunar material. Nature Geoscience 5:251–255.

**SUPPLEMENTARY MATERIALS**

ANALYTICAL PROCEDURES

*Mineralogy*

Several thin sections of Ejby were studied by optical and electron microscopy. For optical microscopy in transmitted and reflected light a ZEISS polarizing microscope (Axiophot) was used. The identification of the different mineral phases in fine-grained components was performed with a JEOL 6610-LV electron microscope (SEM) at the Interdisciplinary Center for Electron Microscopy and Microanalysis (ICEM) at the Westfälische Wilhelms-Universität Münster. Quantitative mineral analyses were obtained using a JEOL JXA-8530F Hyperprobe electron microprobe (EPMA) at ICEM, which was operated at 15 kV and a probe current of 15 nA. Natural and synthetic standards were used for wavelength dispersive spectrometry. Jadeite (Na), kyanite (Al), sanidine (K), chromium oxide (Cr), San Carlos olivine (Mg), hypersthene (Si), diopside (Ca), rhodonite (Mn), rutile (Ti), fayalite (Fe), apatite (P), celestine (S), nickel oxide (Ni), cobalt metal (Co) and tugtupite (Cl) were used as standards for mineral analyses.

*Oxygen isotopes*

Oxygen isotope analysis was carried out at the Open University using an infrared laser-assisted fluorination system (Miller et al. 1999; Greenwood et al. 2017). An approximately 100 mg whole-rock chip of the Ejby chondrite was crushed and homogenized and from this two (~2 mg) replicates were extracted for oxygen isotope analysis (Table 5, Fig. 3). The experimental procedure is described in the supplementary materials section.

Oxygen gas was released from the sample by heating in the presence of $BrF_5$ and this gas was then purified by passing it through two cryogenic nitrogen traps and over a bed of heated KBr. The purified oxygen gas was analysed using a MAT 253 dual inlet mass spectrometer. Overall system precision, as defined by replicate analyses of our internal obsidian standard (n = 38), is: ±0.053 ‰ for $\delta^{17}O$; ±0.095 ‰ for $\delta^{18}O$; ±0.018 ‰ for $\Delta^{17}O$ (2σ) (Starkey et al. 2016).

Oxygen isotopic analyses are reported in standard δ notation, where $\delta^{18}O$ has been calculated as: $\delta^{18}O = [(^{18}O/^{16}O)_{sample}/(^{18}O/^{16}O)_{VSMOW} -1]$ 1000 (‰) and similarly for $\delta^{17}O$ using the $^{17}O/^{16}O$ ratio. $\Delta^{17}O$, which represents the deviation from the terrestrial fractionation line, has been calculated using a linearized format (Miller, 2002):

$$\Delta^{17}O = 1000 \ln(1+ \delta^{17}O/1000) - \lambda\ 1000 \ln(1+ \delta^{18}O/1000)$$

where λ = 0.5247, which was determined using 47 terrestrial whole-rock and mineral separate samples (Miller et al. 1999; Miller, 2002).

*Titanium and Chromium isotope analyses*

The Ti and Cr isotope data were obtained at the Institute of Geochemistry and Petrology at ETH Zurich. The analytical procedure followed that described in Bischoff et al. (2017). A 50 mg and 12.65 mg sample aliquot for Ti and Cr isotope analyses, respectively, were dissolved in a 6 ml square body Savillex Teflon beaker with HF:HNO$_3$ in an oven at 160 ºC for 3-4 days. For Ti purification, a three-stage ion exchange procedure was adapted from Williams (2015) and Schönbächler et al. (2004). The Ti isotope compositions were measured using a Neptune Plus MC-ICPMS in high and medium resolving power mode in conjunction with a Cetac Aridus II desolvating system. The samples were bracketed by an in-house Ti wire standard solution (Williams, 2015). The data was internally normalized to $^{49}$Ti/$^{47}$Ti =0.749766 (Niederer et al. 1981) using the exponential law. The sample uncertainty is reported as standard error based on repeated analyses of the sample solution obtained on three different days. Chromium was separated from the matrix elements employing a two-stage ion exchange procedure adapted from Trinquier et al. (2008). Chromium isotopes were measured on a Neptune Plus MC-ICPMS using a Cetac Aridus II desolvating nebuliser as introduction system. Analyses were performed at medium-resolution (6000<M/ΔM <9000) to resolve molecular isobaric interferences (ArC, ArN, and ArO). Samples and standards were corrected for instrumental mass bias using the exponential law and $^{50}$Cr/$^{52}$Cr = 0.051859 (Shields et al. 1966). The Cr isotope data is reported as the parts per ten thousand deviation from the Cr NIST SRM 979 standard. The sample was measured 3 times and the uncertainty is given as standard error of these repeat measurements.

*Chemistry*

The chemical composition of the bulk sample was obtained using Inductively Coupled Plasma - Atomic Emission Spectroscopy (ICP-AES) and Inductively Coupled Plasma - Sector Field Mass Spectrometry (ICP-SFMS) (Table 2). A 1 g whole-rock sample was analyzed for major and trace element concentrations following the procedures of Barrat et al. (2012, 2016).

*Noble gases*

Noble gases were measured at ETH Zurich with standard procedures (Riebe et al. 2017) in two bulk fragments ("small", 15.41±0.01 mg and "large", 117.82±0.01 mg) of the Herlev fragment (see table 1.). All isotopes of He-Xe were measured in three fractions (He-Ne, Ar, Kr-Xe) in a custom-built mass spectrometer. The gases were extracted in one temperature step at 1700°C in a Mo crucible heated by electron impact. A re-extraction of the large sample at ~1750°C showed negligible gas amounts (≤0.1 % of the total gas for all isotopes), demonstrating the complete gas release during the main step. Blank corrections were small for the light noble gases, <0.5 % for all He, Ne and Ar isotopes apart from $^{36}$Ar for the small sample (3 %) but 20 % and 4 % for $^{84}$Kr, and 8 % and 1.3 % $^{132}$Xe in the small and large samples, respectively. All details of the procedures are given by Riebe et al. (2017).

*Soluble organics chemistry*

The extracts for negative mode electrospray Fourier transform ion cyclotron resonance mass spectrometry (ESI(-)-FT-ICR-MS) analysis were prepared and analyzed in the same conditions as described previously in Schmitt-Kopplin et al. (2010). Briefly, a fresh fragment from the Herlev sample (see Table 1) was first washed with methanol (rapid contact with 1 mL methanol that was subsequently discarded) and immediately crushed in an agate mortar with 0.3 mL of LC/MS grade methanol and further transferred into an Eppendorf tube within an ultrasonic bath for 1 minute. The tube was then centrifuged for 3 min. The supernatant methanol extract was directly used for infusion FT-ICR-MS. Prior the sample extraction, great care was used to clean the agate mortar with solvent in ultrasonic bath. A "blank" sample was produced by following the same extraction procedure without any meteorite fragment, and analyzed before and after the meteorite analysis. No significant mass peaks in the mass range of the meteorite extracts were observed.

In order to fully exploit the advantages of FT-ICR-MS, we routinely control the instrument performance by means of internal calibration on arginine clusters prior to any analysis. Relative m/z errors were usually <100 ppb across a range of 150 < m/z < 1,500. The average mass resolution ranged near 1,000,000 at nominal mass 200, 400,000 at mass 400 and 300,000 at masse 600. The Ejby methanol extracts were measured in negative electrospray ionization modes (ESI(-)) under conditions described earlier (Schmitt-Kopplin et al. 2010); 3000 scans were accumulated with 4 million data points. The conversion of the exact masses into elementary composition is shown in more detail in Tziotis et al. (2011).